\documentclass{iau}
\usepackage{graphicx}
\usepackage{natbib}
\usepackage{epstopdf}
\newcommand{\kms}{~km~s^{-1}}

\title[]
{The intriguing life of star-forming galaxies in the redshift range 1 $\leq$ z $\leq$ 2 using MASSIV}

\author[
	 P. Amram
	\& al.]
{
	P. Amram$^1$,
	C. L\'{o}pez-Sanjuan$^{1,2}$,
	B. Epinat$^1$,
	T. Contini$^3$,
	D. Vergani$^4$,
	L. Tasca$^1$,
	O. Le F\`evre$^1$,
	B. Garilli$^5$,
	C. Divoy$^3$,
	J. Queyrel$^3$,
	M. Kissler-Patig$^{6,7}$,
	J. Moultaka$^3$,
	L. Paioro$^5$,
	L. Tresse$^1$,
	V. Perret$^1$
	\& F. Bournaud$^8$
}

\affiliation{
$^1$LAM, AMU, CNRS, Marseille (F), email: {\tt Philippe.Amram@oamp.fr},
$^2$FEFCA, Teruel (E),
$^3$IRAP, Toulouse (F),
$^4$INAF, Bologna (I),
$^5$IASF-INAF, Milano (I),
$^6$ESO, Garching b. M\"unchen (G),
$^7$GEMINI, Hilo (USA),
$^8$CEA, SAp, AIM, Saclay (F).
}

\pubyear{2013}
\volume{295}  
\pagerange{xx--xx}
\jname{The intriguing life of massive galaxies}
\editors{D. Thomas, A. Pasquali \& I. Ferreras, eds.}
\def\jnl@style{\it}
\def\aaref@jnl#1{{\jnl@style#1}}

\def\aaref@jnl#1{{\jnl@style#1}}

\def\aj{\aaref@jnl{AJ}}                   
\def\araa{\aaref@jnl{ARA\&A}}             
\def\apj{\aaref@jnl{ApJ}}                 
\def\apjl{\aaref@jnl{ApJ}}                
\def\apjs{\aaref@jnl{ApJS}}               
\def\ao{\aaref@jnl{Appl.~Opt.}}           
\def\apss{\aaref@jnl{Ap\&SS}}             
\def\aap{\aaref@jnl{A\&A}}                
\def\aapr{\aaref@jnl{A\&A~Rev.}}          
\def\aaps{\aaref@jnl{A\&AS}}              
\def\azh{\aaref@jnl{AZh}}                 
\def\baas{\aaref@jnl{BAAS}}               
\def\jrasc{\aaref@jnl{JRASC}}             
\def\memras{\aaref@jnl{MmRAS}}            
\def\mnras{\aaref@jnl{MNRAS}}             
\def\pra{\aaref@jnl{Phys.~Rev.~A}}        
\def\prb{\aaref@jnl{Phys.~Rev.~B}}        
\def\prc{\aaref@jnl{Phys.~Rev.~C}}        
\def\prd{\aaref@jnl{Phys.~Rev.~D}}        
\def\pre{\aaref@jnl{Phys.~Rev.~E}}        
\def\prl{\aaref@jnl{Phys.~Rev.~Lett.}}    
\def\pasp{\aaref@jnl{PASP}}               
\def\pasj{\aaref@jnl{PASJ}}               
\def\qjras{\aaref@jnl{QJRAS}}             
\def\skytel{\aaref@jnl{S\&T}}             
\def\solphys{\aaref@jnl{Sol.~Phys.}}      
\def\sovast{\aaref@jnl{Soviet~Ast.}}      
\def\ssr{\aaref@jnl{Space~Sci.~Rev.}}     
\def\zap{\aaref@jnl{ZAp}}                 
\def\nat{\aaref@jnl{Nature}}              
\def\iaucirc{\aaref@jnl{IAU~Circ.}}       
\def\aplett{\aaref@jnl{Astrophys.~Lett.}} 
\def\apspr{\aaref@jnl{Astrophys.~Space~Phys.~Res.}}
\def\bain{\aaref@jnl{Bull.~Astron.~Inst.~Netherlands}}
\def\fcp{\aaref@jnl{Fund.~Cosmic~Phys.}}  
\def\gca{\aaref@jnl{Geochim.~Cosmochim.~Acta}}   
\def\grl{\aaref@jnl{Geophys.~Res.~Lett.}} 
\def\jcp{\aaref@jnl{J.~Chem.~Phys.}}      
\def\jgr{\aaref@jnl{J.~Geophys.~Res.}}    
\def\jqsrt{\aaref@jnl{J.~Quant.~Spec.~Radiat.~Transf.}}
\def\memsai{\aaref@jnl{Mem.~Soc.~Astron.~Italiana}}
\def\nphysa{\aaref@jnl{Nucl.~Phys.~A}}   
\def\physrep{\aaref@jnl{Phys.~Rep.}}   
\def\physscr{\aaref@jnl{Phys.~Scr}}   
\def\planss{\aaref@jnl{Planet.~Space~Sci.}}   
\def\procspie{\aaref@jnl{Proc.~SPIE}}   

\begin{document}

\maketitle

\begin{abstract}

MASSIV (Massiv Assembly Survey with SINFONI in VVDS) is an ESO large program which consists of 84 star-forming galaxies, spanning in a wide range of stellar masses, observed with the IFU SINFONI on the VLT, in the redshift range $1 \leq z \leq 2$.  To be representative of the normal galaxy population, the sample has been selected from a well-defined, complete and representative parent sample. The kinematics of individual galaxies reveals that $58\%$ of the galaxies are slow rotators, which means that a high fraction of these galaxies should probably be formed through major merger processes which might have produced  gaseous thick or spheroidal structures supported by velocity dispersion rather than by rotation. Computations on the major merger rate from close pairs indicate that a typical star-forming galaxy underwent $\sim0.4$ major mergers since $\sim9.5~Gyr$, showing that merging is a major process driving mass assembly into the red sequence galaxies. These objects are also intriguing due to the fact that more than one galaxy over four is more metal-rich in its outskirts than in its center.

\end{abstract}

\firstsection 

\section{Mass assembly of high redshift galaxies}

Processes of galaxy mass assembly at early epoch is one amongst the largest open issues in galaxy evolution.
To tackle this question several mechanisms related to the environment should be first understood, e.g. the role of major and/or minor, wet and/or dry mergers versus smooth cold gas accretion along cosmic filaments. On the other hand, the actual impact of feedback from SNe and AGNs activities and the secular galaxy evolution should also play a major role in the way the baryonic matter is redistributed within - or ejected from - the galaxies.  The knowledge of numerous physical parameters is needed to constrain scenarios of galaxy mass assembly.  Number of global parameters could be achieved from global measurements but resolved measurements within individual galaxies extracted from 3D-spectral analysis are necessary to measure the radial distributions of the metallicity, the star formation, the stellar and gaseous masses. With the present-day observational facilities, resolved line measurements could not yet be achieved for galaxies at redshifts higher than $z>4$. In the redshift range $2 \leq z \leq 4$, because of strong biases due to color pre-selections, only the brightest star-forming galaxies could be studied which is not the case in the redshift range $1 \leq z \leq 2$, the 2.6 Gyr-lasting epoch where the stark dichotomy within the galaxy population is rising and is set at $z\sim1$ and when star-forming galaxies representative of the bulk of galaxy population could be addressed.

MASSIV (Massiv Assembly Survey with SINFONI in VVDS) is an ESO large program which consists of 84 star-forming galaxies, in the redshift range $1 \leq z \leq 2$, observed from 2008 to completion in 2011, with the NIR-IFU SINFONI on the VLT. Spanning in a wide range of stellar masses log(M*)=[9,12] and selected from a well-defined, complete and representative parent sample based on accurate spectroscopic  redshifts from the VIMOS VLT Deep Survey \citep{LeFevre:2005}, the MASSIV sample is representative of the normal galaxy population. The J- or H-band has been used to target the redshifted $H\alpha$ bright emission line with a high spatial resolution ($<0.8''$), partly with adaptative optics. A full survey description and global properties of the galaxy sample is given in \citeauthor{contini:2012} (\citeyear{contini:2012}, Paper I). The first epoch sample (50/84 galaxies) leads to several studies about kinematics and close environment classification (\citealp{epinat:2012}, Paper II); evidence for galaxies displaying positive metallicity gradients (\citealp{queyrel:2012}, Paper III); fundamental relations (\citealp{vergani:2012}, Paper IV) and major merger rate from close pairs (\citealp{Lopez-Sanjuan:2012}, Paper V).


\section{Do high-redshift star-forming galaxies rotate ?}

The total MASSIV sample of 84 galaxies allows resolved velocity measurements for 76 galaxies. For the kinematic classification, we used several criteria \citep{epinat:2012}. One of them is based on the total velocity shear $V_{shear}$ measured on the velocity field not inclination-corrected: 32/76 (42\%) galaxies have a high velocity shear ($V_{shear}> 100 \kms$) and 44/76 (58\%) a low velocity shear ($V_{shear}<100 \kms$). Another criterion is to distinguish rotators from non-rotating galaxies using a classification based on the disagreement between morphological and kinematic position angles and the mean weighted velocity field residuals normalized by the velocity shear: again 32/76 (42\%) galaxies are rotators while 44/76 (58\%) are not rotating objects. Finally we consider the nearby environment to distinguish isolated from not isolated galaxies: 58/76 (76\%) galaxies are isolated while 18/76 (24\%) are not. The galaxies presenting large $V_{shear}$ are mainly rotating disks but also interacting/merging systems. Galaxies showing low $V_{shear}$ may be face-on disks, star-forming spheroids or on-going mergers in transient state. For rotating disks, depending on the total mass of the galaxies, typical maximum rotation velocities $V_{rot}$ range between $100$ to $300\kms$. Considering these typical maximum rotation velocities $V_{rot}=200^{+100}_{-100}\kms$, if all the disks were pure rotators in planes randomly distributed in the sky, the number of low $V_{shear}$ disks should not exceed $11$ galaxies $(14\%)$ if all the galaxies were low mass galaxies (for $V_{rot}=100~\kms$), $3$ galaxies $(4\%)$ for $V_{rot}=200~\kms$ and $1.5$ galaxies ($2\%$) if all the galaxies were high mass galaxies (for $V_{rot}=300~\kms$). Furthermore, even if the actual galaxy masses are overestimated, the high fraction of $58\%$ of low velocity shear is incompatible with the hypothesis of face-on rotators.

Finally, it has been shown in \cite{vergani:2012} that \textit{(1)} non-rotating galaxies are more compact in their extent of the stellar component than rotators but they are not statistically different in their gas extent; \textit{(2)} marginal evolution in the size-stellar mass and size-velocity relations are observed and \textit{(3)} the large dispersion observed in the stellar and baryonic Tully-Fisher relations is reduced using the $S_{0.5}=\sqrt(0.5\times V_{rot}^2+\sigma_0^2)$ index instead of $V_{rot}$ ($\sigma_0$ being the velocity dispersion). Nevertheless, an intrinsic spread around the median trend remains even when using the $S_{0.5}$ index instead of $V_{rot}$ and slowly rotating galaxies display lower $S_{0.5}$ index and stellar/baryonic masses than fast rotators.

\section{One star-forming galaxy over four is more metal-rich in its outskirts than in its centers}

Results on the metallicity gradients for the first epoch sample (50 galaxies) have been published in \cite{queyrel:2012} while the second epoch sample (35 galaxies) is still under analysis (Divoy et al., in preparation). Among the $50$ galaxies of the first epoch, both $H\alpha$ and $[NII]$ lines have been measured for $34$ galaxies and metallicity gradients for $26/34$. Positive (negative) gradients refers to metallicity being higher (lower) in the center than in the outskirts. We have measured secure negative gradients for $5/26$ galaxies and positive ones for $7/26$. Preliminary results on the second epoch sample, which contains in average fainter and smaller galaxies than the first epoch one, indicate that integrated metallicities are measurable for $\sim20/35$ galaxies.  Metallicity gradients have been measured for $10/20$ galaxies and forthcoming measurements will probably be possible for an extra couple of galaxies.  Among those 10 galaxies, $7$ of them display negative gradients and $3$ positive ones. On the whole MASSIV sample presently available, metallicity gradients have been measured on $36$ galaxies; $12$ galaxies ($33\%$) show secure classical negative gradients but $10$ galaxies ($28\%$) show unexpected positive gradients, the other $14$ galaxies ($39\%$) are compatible with a flat radial metallicity distribution. $9$ galaxies over $12$ ($75\%$) showing a negative gradient are isolated ($3/12$ or $25\%$ are in interaction) while $6$ galaxies over $10$ ($60\%$) that show a positive gradient are interacting ($4/10$ or $40\%$ are isolated).  Interactions, mergers or cold gas accretion might be responsible for shallowing and even inverting the abundance gradient.

\section{Star-forming galaxy underwent $\sim0.4$ major mergers since $\sim9.5~Gyr$}

\begin{figure}[b]
\begin{center}
 \includegraphics[width=6.7cm]{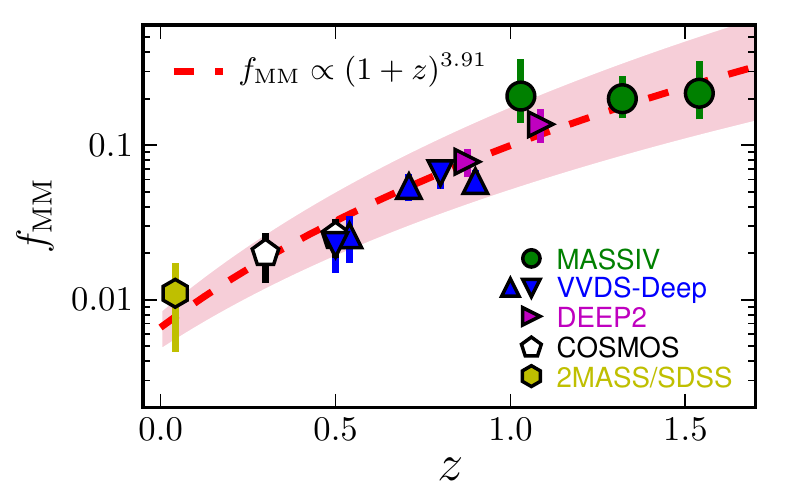}\hfill
 \includegraphics[width=6.7cm]{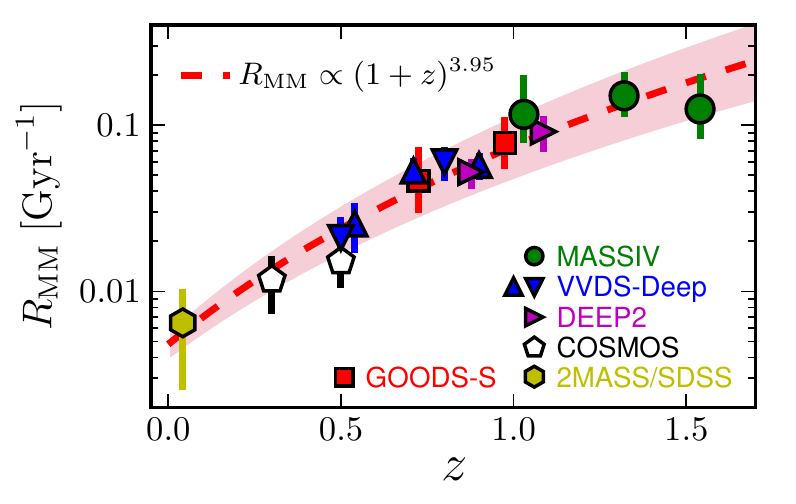}
 \caption{Major merger fraction (left panel) and Major merger rate (right panel) of $M_* \sim 10^{10-10.5} M_{\odot}$ galaxies as a function of redshift. On both panels, diamonds are from this MASSIV data set, triangles from \cite{deravel:2009} and inverted triangles from \cite{lopez-sanjuan:2011}, both in VVDS-Deep, pentagons are from \cite{Xu:2012} in the COSMOS field, and the hexagon is from \cite{Xu:2012} in 2MASS/SDSS. On the right panel only, squares are from \cite{lopez-sanjuan:2009} in GOODS-S, pentagons are from \cite{Xu:2012} in the COSMOS field. On both panels, the solid line is the error-weighted least-squares fit of a power-law function, respectively $f_{MM} = 0.0062\times(1+z)^{3.9}$ and $R_{MM} = 0.0058\times(1+z)^{3.7}$, to the data. The grey areas mark the 68\% confidence interval in the fit.}
   \label{fig1}
\end{center}
\end{figure}

The major merger rate at $0.9 \leq z \leq 1.7$ from IFU-based close pairs in MASSIV is studied in \cite{Lopez-Sanjuan:2012}. A close pair is defined by a couple of galaxies showing \textit{(1)} a projected separation lower than $20~h^{-1}kpc$ and \textit{(2)} a radial velocity difference lower than $500\kms$. When the two galaxies overlap, the two components have been separated both using the morphology and the kinematics (see left panels of Fig. \ref{fig1}). A major (minor) close pair is defined by a couple of galaxies for which the luminosity ratio is $L_2/L_1>1/4$ ($L_2/L_1<1/4$). $N_P$ is the number of major close pairs over a population of $N$ principal galaxies targeted. The fraction of major merger $f_{MM}$ is basically equal to $N_P/N$ corrected for selection effects and covered area. Within the stellar mass range $10^9-10^{10} M_{\odot}$, the MASSIV sample contains 20 close pair candidates, including $N_P=13$ major mergers and 7 minors mergers. The major merger rate $R_{MM}\propto f_{MM}/T_{MM}$ at three different epochs, where $T_{MM}$ is the merger time scale extracted from the Millennium simulation, is given in table \ref{tab1}. Using the MASSIV data combined with data extracted for the literature, Fig.~\ref{fig1} shows that the fraction of major mergers evolves with the redshift $z$ like $f_{MM}\propto(1+z)^{3.91}$ and the major merger rate like $R_{MM}\propto(1+z)^{3.95}$. The average number of major gas-rich mergers per star-forming galaxy between $z=1.5$ and $z=0$ is $0.37^{+0.21}_{-0.13}$.  Half of the merger activity occurs at high redshift between $z=1.5$ and $z=1$ ($\sim 1.6~Gyr$) and the other half more recently between $z=1.0$ and $z=0$ ($\sim 7.7~Gyr$) which means the average merger activity was higher by a factor $5$ at the earliest epoch than at the later one. Accordingly to early type galaxies (ETGs) classification based on fast and slow rotators from \cite{Emsellem:2011} and using the following assumptions: \textit{(1)} wet major mergers produce fast rotators; \textit{(2)} dry major mergers produce slow rotators, \textit{(3)} dry minor mergers do not change the kinematical state of ETGs and \textit{(4)} all ETGs at z=1.3 are fast rotators, a fraction of slow rotator of $55\%$ is computed and lead to the conclusion that merging is a major process driving mass assembly into the red sequence galaxies. The right panel of Fig. \ref{fig1} shows that the combined effect of gas-rich and dry mergers  $f_{tot} = f_{wet} + f_{dry}$ is able to explain the evolution in ETGs since $z\sim1.3$, with gas rich merging accounting for $2/3$. Minor merging is definitively present in the MASSIV sample but due to incompleteness we are not able to assess a minor merger rate.

\begin{table}
  \begin{center}
  \caption{Fraction of major merger $f_{MM}$ and major merger rate $R_{MM}$}
  \label{tab1}
    \begin{tabular}{lcc}\hline
    $T_{MM}~(Gyr)~[z_1,z_2]$ & $f_{MM}~(\%)~@~z~$ & $R_{MM}~(Gyr^{-1})~@~z~$\\
    \hline
    $1.80~[0.94,1.06]$ & $21~@~z~=1.0$& $0.12~@~z~=1.0$\\
    $1.37~[0.20,1.50]$ & $20~@~z~=1.4$& $0.15~@~z~=1.4$\\
    $2.54~[1.50,1.80]$ & $23~@~z~=1.6$& $0.13~@~z~=1.6$\\
    \hline
    \end{tabular}
  \end{center}
\end{table}

\begin{figure}[b]
\begin{center}
 \includegraphics[width=4.1cm]{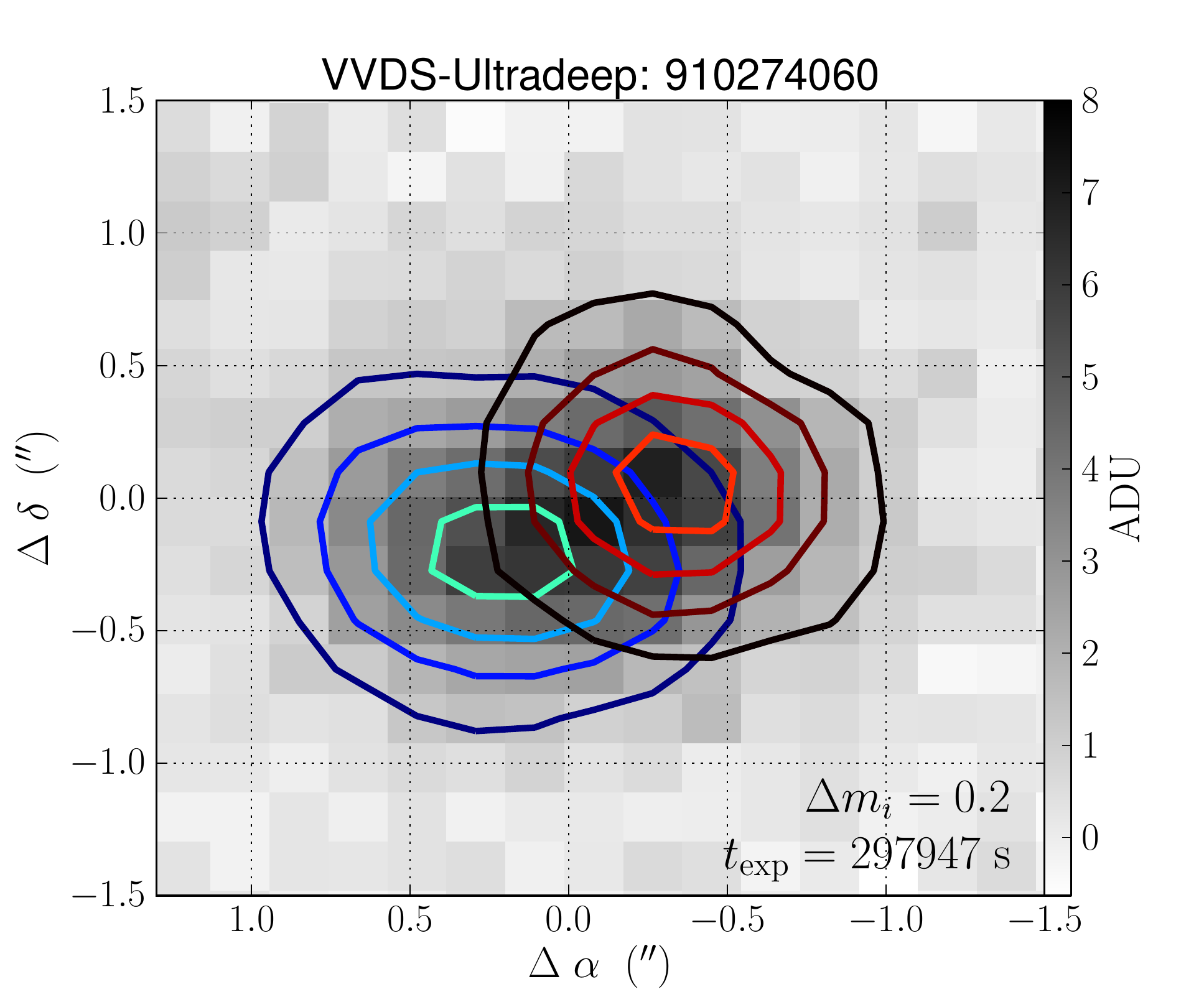}\hfill
 \includegraphics[width=4.1cm]{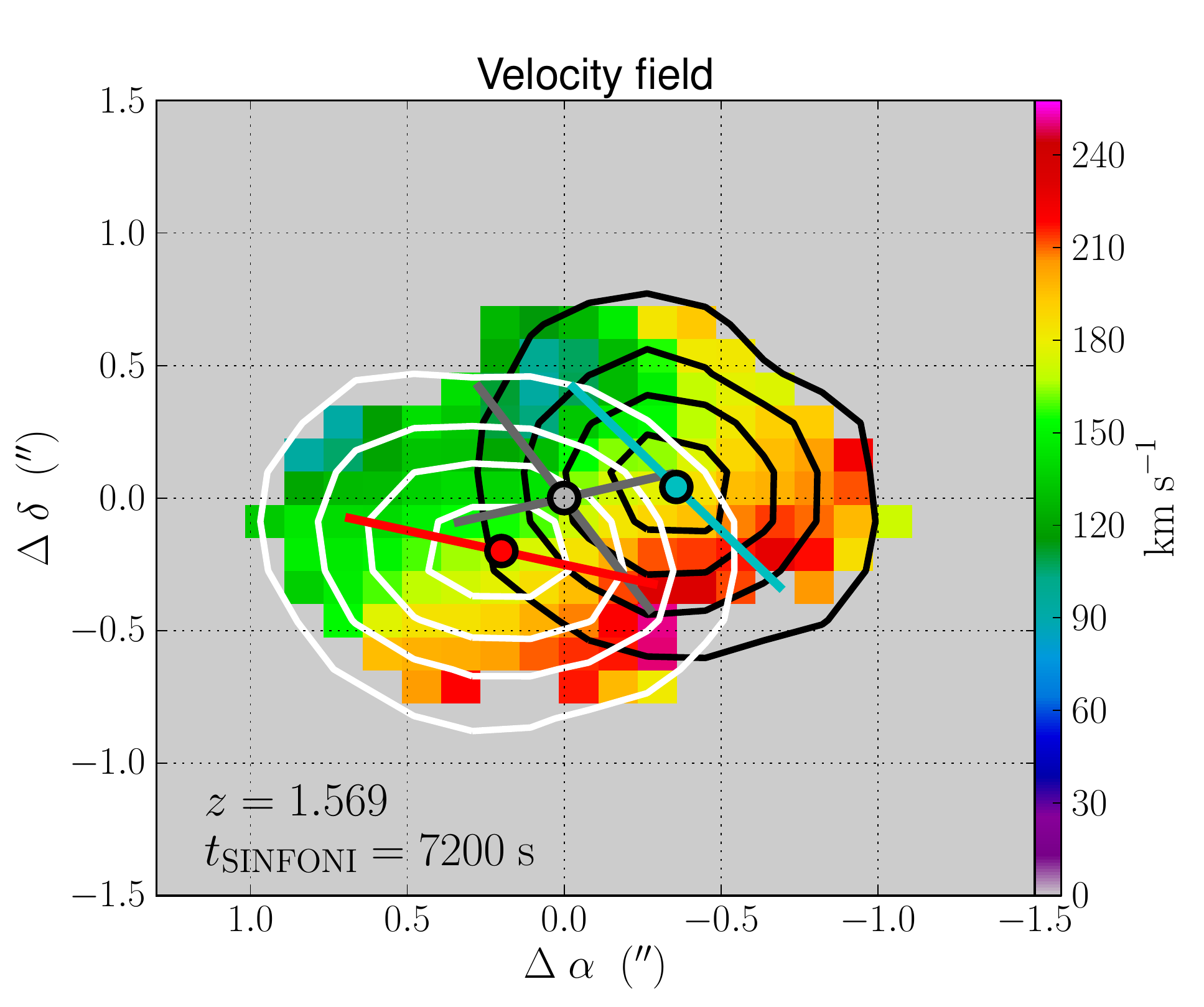}\hfill
 \includegraphics[width=5.2cm]{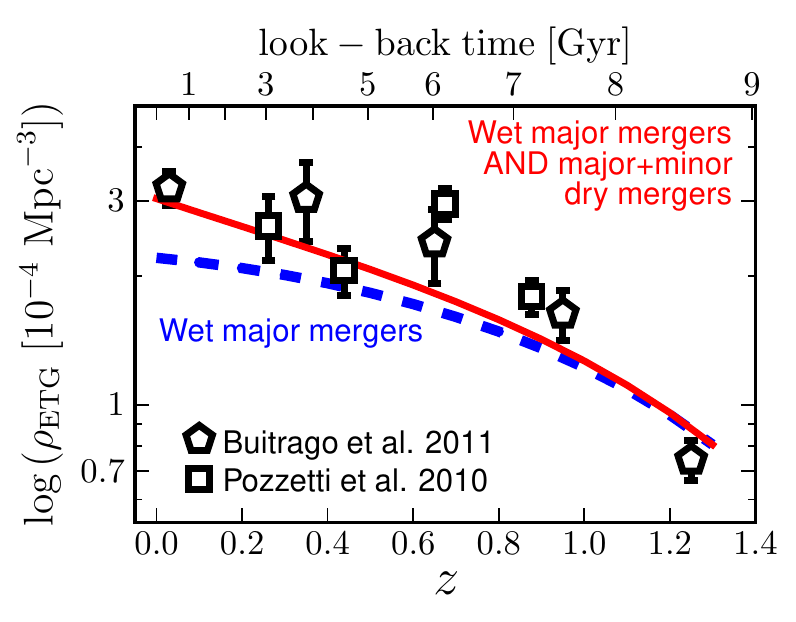}
 \caption{i-band image (left panel) and Velocity field (middle panel) of the MASSIV source 910274060 (major merger). North is up and East is left. The level contours mark the isophotes of the two components obtained with GALFIT in the i-band image. The principal galaxy (red/white) is the one closer to the kinematical centre of the system while the companion (blue/black) is the secondary component. The outer contour was chosen to fit the kinematical maps. (Right panel) Number density evolution of massive ($M_* \geq 10^{11.25} M_{\odot}$) ETGs (E/S0) as a function of redshift from \cite{Buitrago:2011}(pentagons).}
   \label{fig2}
\end{center}
\end{figure}

\end{document}